\documentclass{nature}
\usepackage{graphicx}
\usepackage{subfigure}
\usepackage{dcolumn}
\usepackage{bm}
\usepackage[mathlines]{lineno}
\usepackage{array}
\usepackage{braket}
\usepackage{diagbox}
\usepackage{amsmath}
\usepackage{appendix}
\usepackage{multirow}
\usepackage{bm,color,bbm}
\usepackage{float}
\newcolumntype{.}{D{.}{.}{-1}}
\usepackage{amsopn}
\usepackage[colorlinks, linkcolor=red, anchorcolor=black, urlcolor=red, citecolor=blue]{hyperref}
\usepackage{epstopdf}

\newcommand{\blk}{\color{black}}

\begin{document}
\title{High-speed measurement-device-independent quantum key distribution with integrated silicon photonics}
\author
{Kejin Wei$^{1,2,3,\ast}$, Wei Li$^{1,2,\ast}$, Hao Tan$^{1,2}$, Yang Li$^{1,2}$, Hao Min$^{1,2}$, Wei-Jun Zhang$^{4}$, Hao Li$^{4}$, Lixing You$^{4}$, Zhen Wang$^{4}$, Xiao Jiang$^{1,2}$, Teng-Yun Chen$^{1,2}$, Sheng-Kai Liao$^{1,2}$, Cheng-Zhi Peng$^{1,2}$, Feihu Xu$^{1,2,\star}$, Jian-Wei Pan$^{1,2,\star}$}

\maketitle

\begin{affiliations}
\item National Laboratory for Physical Sciences at Microscale and Department of Modern Physics, University of Science and Technology of China, Hefei 230026, China.
\item CAS Center for Excellence and Synergetic Innovation Center in Quantum Information and Quantum Physics, University of Science and Technology of China, Hefei 230026, China
\item Guangxi Key Laboratory for Relativistic Astrophysics, School of Physics Science and Technology, Guangxi University, Nanning 530004, China
\item State Key Laboratory of Functional Materials for Informatics, Shanghai Institute of Microsystem and Information Technology, Chinese Academy of Sciences, Shanghai 200050, China \\
$^\ast$These authors contributed equally.
\\$^\star$e-mails: feihuxu@ustc.edu.cn; pan@ustc.edu.cn
\end{affiliations}

\begin{abstract}
Measurement-device-independent quantum key distribution (MDI-QKD)\cite{lo2012measurement} removes all detector side channels and enables secure QKD with an untrusted relay. It is suitable for building a star-type quantum access network\cite{2013Frohli,hughes2013network}, where the complicated and expensive measurement devices are placed in the central untrusted relay and each user requires only a low-cost transmitter, such as an integrated photonic chip\cite{2017sibson,2016Ma,2016Sibson}. Here, we experimentally demonstrate a 1.25 GHz silicon photonic chip-based MDI-QKD system using polarization encoding. The photonic chip transmitters integrate the necessary encoding components for a standard QKD source. We implement random modulations of polarization states and decoy intensities, and demonstrate a finite-key secret rate of 31 bps over 36 dB channel loss (or 180 km standard fiber). This key rate is higher than state-of-the-art MDI-QKD experiments. The results show that silicon photonic chip-based MDI-QKD, benefiting from miniaturization, low-cost manufacture and compatibility with CMOS microelectronics, is a promising solution for future quantum secure networks.
\end{abstract}

\maketitle

\paragraph{Introduction.}

\begin{figure}[!hbt]
	\centering
	\includegraphics[width=0.7\linewidth]{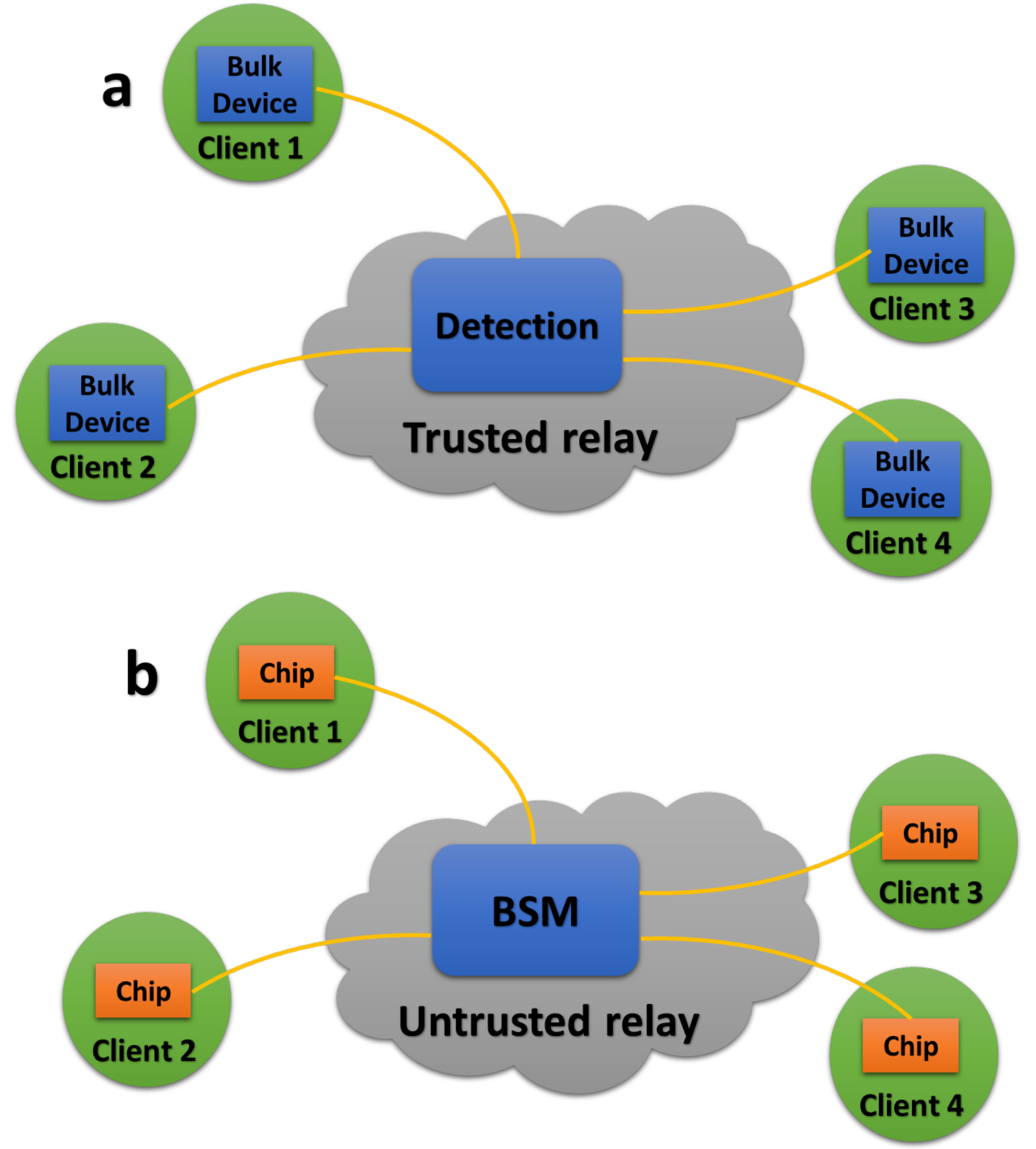}
\caption{\textbf{a,} Schematic of a quantum access network with detectors as the shared resource using multiplexing device, such as a dense-wavelength-division multiplexing. The central relay needs to be trustful where the detectors are vulnerable to quantum attacks. \textbf{b,} Schematic of a star-type chip-based MDI-QKD network. Each client holds only a QKD transmitter chip, and the central relay performs the Bell state measurement (BSM) to remove all detector attacks, enabling secure QKD with an untrusted relay.
		\label{network}}
\end{figure}

QKD\cite{2014Lo} is a key technology for building nodal networks which are believed to be a crucial stepping stone towards a quantum internet. So far, existing QKD networks\cite{2009Peev,2010Chen,2011Sasaki,2013Frohli,hughes2013network} need the central relays to be trusted (e.g., Fig. \ref{network}a), which is a critical security drawback\cite{2014Lo}. Fortunately, the MDI-QKD protocol\cite{lo2012measurement} (see also ref.\cite{2012Braunstein}) can remove all side channels of the measurement devices\cite{2010lydersen}, and it is practical with current technology. MDI-QKD has been widely implemented towards long distance\cite{2014yanlin,2016Yin}, high secret key rate\cite{2016Comandar}, field test\cite{2013Ruben,2016Tangnetwork} and so forth\cite{2016zhiyuan,2017vali,wang2017measurement,liu2018experimental}. See ref.\cite{2019Xu} for a review. Recently, a MDI-typed scheme, twin-field QKD\cite{2018Lucamarini}, was proposed to overcome the repeaterless key-rate bound.

Chip-based QKD has attracted great attention\cite{2017sibson,2016Ma,2016Sibson,bunandar2018metropolitan,paraiso2019modulator,ding2017high,raffaelli2018homodyne,zhang2019integrated,avesani2019full}, due to its advantages of compact size and low cost. Particularly, silicon that relies on well-established
fabrication techniques is well suited for on-chip photonic QKD components, and it has been exploited to implement several QKD protocols, including decoy-state BB84\cite{2016Ma,2016Sibson,bunandar2018metropolitan,paraiso2019modulator}, high dimension\cite{ding2017high}, continuous variable\cite{raffaelli2018homodyne,zhang2019integrated} and so forth\cite{2017sibson,avesani2019full}.

The combination of silicon photonic chips and MDI-QKD enables a remarkably new network-centric\cite{hughes2013network} or quantum-access\cite{2013Frohli} structure with an \emph{untrusted} relay. In such a structure (see Fig.~\ref{network}b), each user only needs a compact transmitter chip, whereas the relay holds the expensive and bulky measurement system (and quantum memory\cite{bhaskar2019experimental}) which are shared by all users. Importantly, this structure can by-pass the challenging technique for intergrading single-photon detectors on chip\cite{pernice2012high,najafi2015chip}, since the users do not need to do the quantum detection. Overall, the chip-based MDI-QKD network is a promising solution for low-cost, scalable QKD networks with an untrusted relay.

Here, we experimentally demonstrate a 1.25 GHz, silicon-chip-based, polarization-encoding MDI-QKD system. Each user possesses a photonic chip transmitter, which integrates the QKD encoding components of intensity modulator, polarization modulator and variable optical attenuator. The chips are manufactured by standard Si photonic platforms, packaged with thermoelectric cooler (TEC), and designed compactly for the purpose of commercial production. With two chip transmitters, we implement MDI-QKD with random modulations of decoy intensities and polarization qubits, and demonstrate a finite-key secret rate of 31 bps over 36 dB channel loss. In addition, we obtain a key rate of 497 bps over 140 km commercial fibre spools. The achieved key rate is higher than those of previous MDI-QKD experiments\cite{2014yanlin,2016Yin,2016Comandar,2013Ruben,2016Tangnetwork,2016zhiyuan,2017vali} (see Table \ref{Comparison}).

\begin{figure*}[!hbt]
	\centering
	\includegraphics[width=1\linewidth]{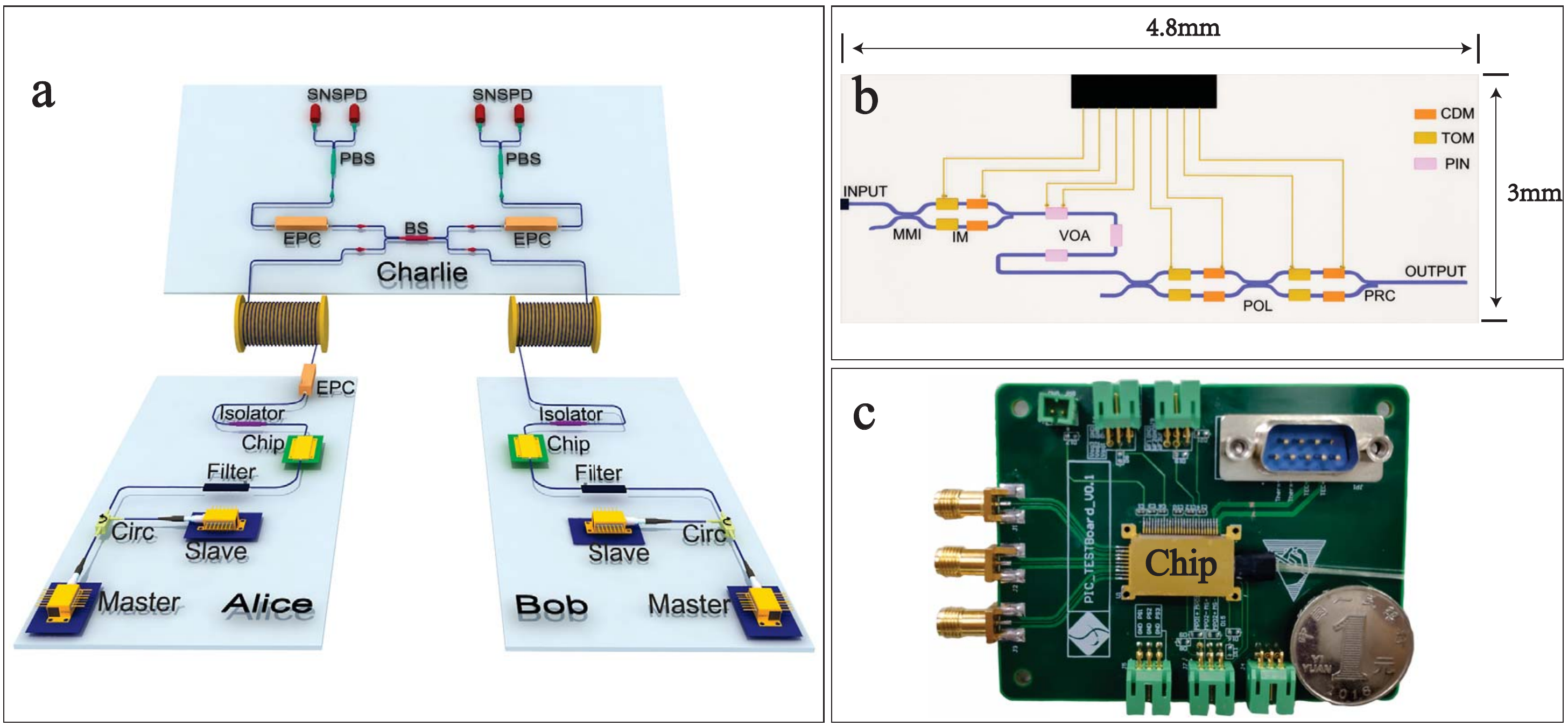}
	\caption{\textbf{a}, Experimental set-up of chip-based MDI-QKD. Alice and Bob each possess a master gain-switched laser (Master) injects photons into the cavity of a slave gain-switched laser (Slave) through a circulator (Circ) to generate low-jitter phase-randomized light pulses at a repetition rate of 1.25~GHz. The generated pulses are coupled into a silicon photonic transmitter chip (Chip) which integrated together an intensity modulator, variable optical attenuator, and polarization modulator. The Bell state measurements are performed by the untrusted relay Charlie who comprises a beam splitter (BS), two electric polarizing controllers (EPCs), two polarizing beam splitters (PBSs) and four superconducting nanowire single photon detectors (SNSPDs).  \textbf{b}, The schematic of the Si Chip. All components are fabricated using standard Si blocks, including multi-mode interference (MMI) couplers, thermo-optics modulators (TOMs),  carrier-depletion modulators (CDMs) and polarization rotator combiner (PRC).  \textbf{c}, Image of the packaged chip soldered to a compact control board.
		\label{setup}}
\end{figure*}

\paragraph{Setup.} Figure~\ref{setup}a shows the schematic of our chip-based MDI-QKD experiment. Using pulsed laser seeding technology\cite{2014Yuan} where a master gain-switched laser (Master) injects photons into the cavity of a slave gain-switched laser (Slave) through a circulator (Circ), Alice and Bob each generates low-jitter phase-randomized light pulses at a repetition rate of 1.25~GHz and a center wavelength of 1550~nm. The generated pulses are pass through a 10~GHz bandwidth filter to remove noise. With these sources, we observe stable Hong-Ou-Mandel interference with a visibility up to 48.4\% (See Methods).

The generated pulses are coupled into a Si photonic transmitter chip which integrates together an intensity modulator, a polarization modulator and a variable optical attenuator. The components are realized by an in-house design\cite{company} comprising several interferometric structures (see Fig. \ref{setup}b) which exploit standard building blocks offered by the IMEC foundry. The multi-mode interference (MMI) couplers act as symmetric beam splitters, and the thermo-optics modulators (TOMs) with $\sim$KHz bandwidth, and carrier-depletion modulators (CDMs) with $\sim$GHz bandwidth act as phase modulators. Specifically, the intensity modulator, which is used to generate decoy state with different intensities, is realized by the first Mach-Zehnder interferometer (MZI) containing both TOMs and CDMs. The next components is the VOA, consisting of a p-i-n (PIN) diode for current injection across-section of the Si waveguide and being used to attenuated the pulses to single-photon levels.  The tunable attenuation is controlled by applying differential biased voltage to the TOMs with an attenuation up to 110~dB. The output of VOA is connected to the polarization modulator (POL) which is realized by combining an inner MZI with two external CDMs ending in polarization rotator combiner (PRC). The POL can prepare the four BB84 states, $\left| \psi  \right\rangle  =( \left| H \right\rangle  + {e^{i\theta }}\left| V \right\rangle)/\sqrt{2} ,\theta  \in \{ 0,{\pi  \mathord{\left/
		{\vphantom {\pi  2}} \right.
		\kern-\nulldelimiterspace} 2},\pi ,{{3\pi } \mathord{\left/
		{\vphantom {{3\pi } 2}} \right.
		\kern-\nulldelimiterspace} 2}\} $. Further details involving with the principle of the design can be found in Methods.

\begin{figure}[!hbt]
	\centering
	\includegraphics[width=0.7\linewidth]{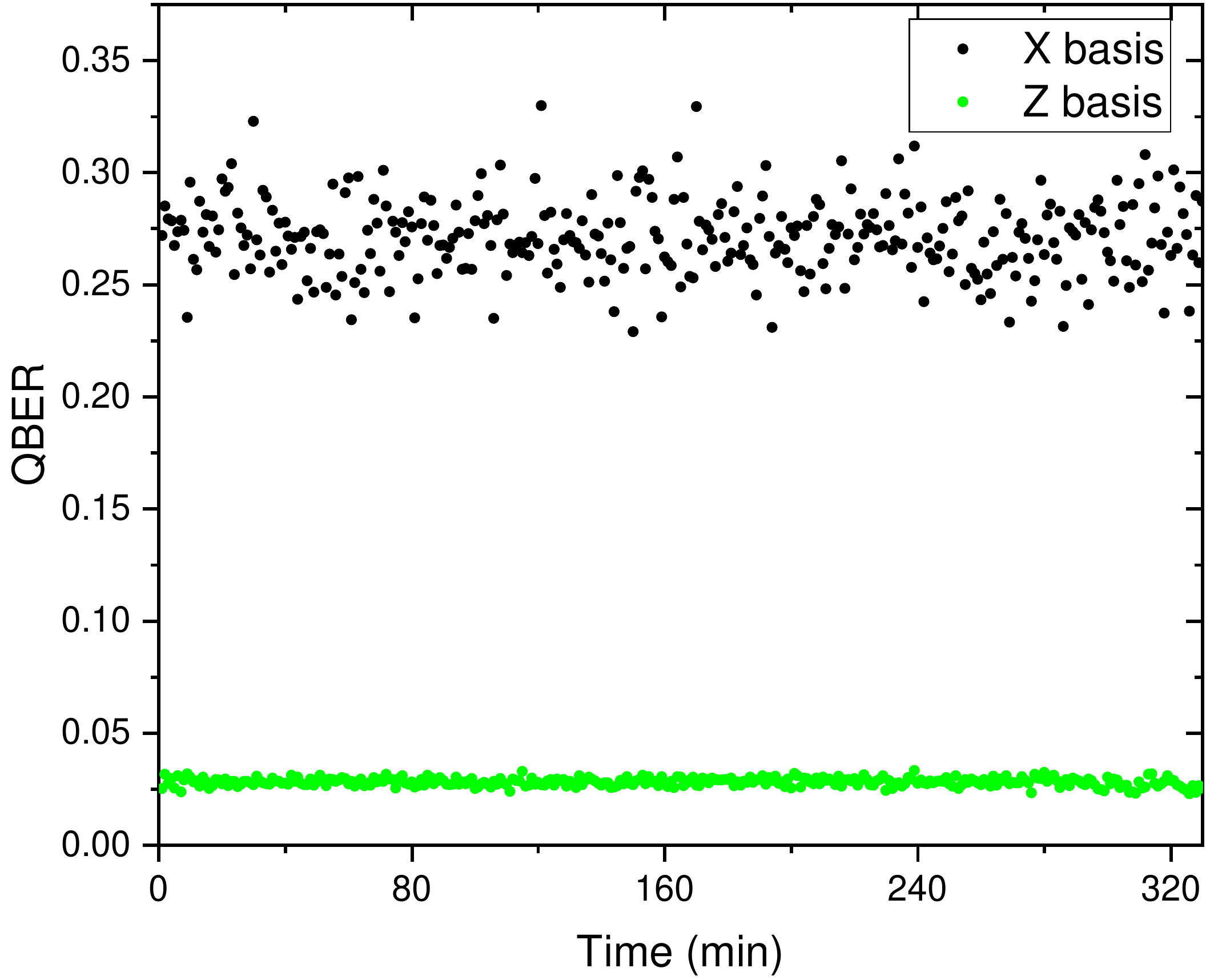}
	\caption{Observed QBERs in different bases over 36 dB channel loss. The data points are collected without any active adjustment of the system. Each point is calculated by using data collected over one-minute period.  An average QBER of 0.028 (0.271) in $Z$ ($X$) basis is observed over several hours of operations.
		\label{Error}}
\end{figure}

The chip has a footprint size of $4.8\times3~mm^2$ and is packaged with a commercial TEC. A precision and compact temperature controller is designed to drive the TEC. With this design, the chip provides a stable polarization encoding and decoy-state modulation. The observed quantum bit error rate (QBER) maintains at low values over several hours of operation (See Fig.~\ref{Error}). The packaged chip with a total volume of $20\times11\times5~mm^3$ is soldered to a standard $9\times7~cm^2$ printed circuit board, as shown in Fig. \ref{setup}c. With dedicated layout, the chip is easily assembled by using commercial foundry, providing a low-cost, portable, stable and miniaturized device for MDI-QKD.

To realize MDI-QKD, Alice and Bob send their encoding pulses to Charlie, who performs a BSM on the incoming pluses. Charlie's measurement setup comprises a 50/50 beam splitter~(BS), two electronics polarisation controllers (EPCs), two polarising beam splitters and four superconducting nanowire single photon detectors (SNSPDs, detection efficiency $\sim$53\%, dark counts $\sim$50~Hz). The detection events are registered using a high-speed time tagger where a successful coincidence induces a projection into one of the two Bell states $\left| {{\psi ^ \pm }} \right\rangle  = {1 \mathord{\left/
		{\vphantom {1 {\sqrt 2 (}}} \right.
		\kern-\nulldelimiterspace} {\sqrt 2 (}}\left| {HV} \right\rangle  \pm \left| {HV} \right\rangle ) $.

For high-rate MDI-QKD, an important part is the high-speed electronics for control and synchronization. In our experiment, Alice and Bob each uses a home-made cost-effective FPGA board to accomplish all electrical controls, including driving the laser, randomly modulating IMs and POLs, synchronizing all stations, etc. The specialised electronics enable us to take advantage of the small size of the chips towards a compact MDI-QKD system. To share a common polarisation reference between Alice and Bob, as well as compensate the polarisation drift in the quantum channel, we develop an automatic polarisation alignment with electronic polarisation controllers, which can rapidly calibrate the polarisation reference (see Supplementary).


\paragraph{Results.} We experimentally characterise each of components in the chip. The bandwidth of the CDM reaches $\sim$21~GHz which is measured by using a vector network analyzer. The IM provides a static extinction ratio (ER) of $\sim$30 dB and a dynamic ER of $\sim$20 dB. We characterize the produced polarisation state with  measurement devices in Charlie. The EPCs are adjusted so that each PBS is aligned to rectilinear and diagonal bases, respectively. We obtain an average polarization ER of $\sim$23 dB. The attenuation of the VOA is ranging from 0 to $\sim$110 dB. The performance of the chip is sufficient for a low-error, high-rate MDI-QKD (see Supplementary).

\begin{figure}[!hbt]
	\centering
	\includegraphics[width=0.7\linewidth]{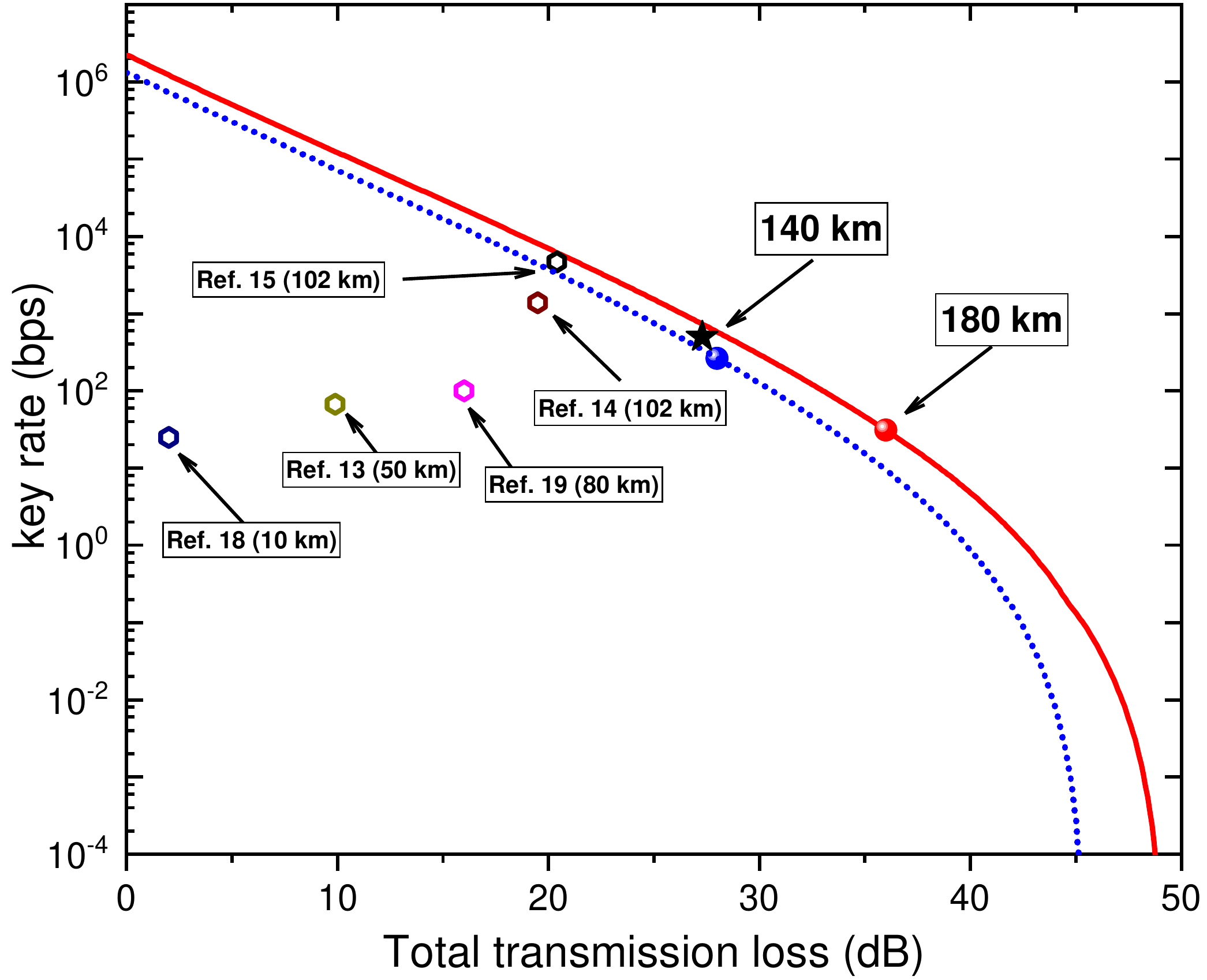}
	\caption{Secure key rates with different transmission loss. The blue and red points show the experimental results with a total transmission loss of 28 dB and 36 dB, respectively. The black star is the key rate obtained using two 70-km commercial fibre spools. The red solid lines and the blue dot lines are theoretical simulations tailored to the corresponding experimental conditions. We also plot the highest finite key rates of current MDI-QKD experiments with polarization encoding (navy diamond\cite{2016zhiyuan} and black diamond\cite{2016Comandar}) and time-bin encoding (dark yellow diamond\cite{2014yanlin}, magenta diamond\cite{2017vali} and brown diamond\cite{2016Yin}).
		\label{keyrate}}
\end{figure}

\blk Using the described set-up, we perform a series of MDI-QKD experiments using the four-intensity decoy-state protocol\cite{2016Zhou}. Finite-key effects are carefully addressed using the standard error analysis approach\cite{xu2014protocol}. In the finite-key scenario\cite{curty2014finite} with a failure probability of $10^{-10}$, we perform a full optimization of the implementation parameters by exploiting the joint constrains for statistical fluctuations\cite{2016Zhou} (see Supplementary). The experimental results are plotted in Fig.~\ref{Error} and Fig.~\ref{keyrate}. The data points are first collected by using optical attenuators to emulate the attenuation of standard single mode fibres (0.2~dB/km). We obtain an average QBER of $\sim$2.8\% (27.1\%) in $Z$ ($X$) basis. At the total loss of 28~dB (corresponding to 140~km fibre), we run the system for 7.7 hours and send a total of $3\times{10^{13}}$ pulse pairs from each client. The finite-key secret rate is 268~bps. At the total loss of 36~dB (corresponding to 180~km fibre), to maximize the key rate, we slightly enhance the bias current of SNSPDs, resulting in a higher detection efficiency (62\%)  but a lower maximum counting rate. We achieve a finite-key secret rate of 31 bps in 10 hours of system operation time. Next, we replace the optical attenuators with two commercial fibre spools of 70 km each (corresponding to $\sim$27~dB tota loss), and obtain an asymptotic secret key rate of 497 bps which is close to the finite-key one obtained from the optical attenuators (see Supplementary).

To illustrate the progress entailed by our results, we include in Fig.~\ref{keyrate} the highest key rate of selection of existing MDI-QKD experiments. See Table~\ref{Comparison} for a detail comparison of different parameters. Although a GHz MDI-QKD was reported in ref.\cite{2016Comandar}, the implementation of random modulations of decoy intensities and polarization states was not demonstrated there. In this sense, apart from the chip-based implementation, our experiment is the first GHz MDI-QKD with random modulations, and also, it represents the highest key rate for MDI-QKD.


\paragraph{Discussions.} We have demonstrated a high-speed chip-based MDI-QKD system where both clients possess a low-cost Si photonic transmitter chip. The transmitter can be further integrated with the laser on a monolithic chip\cite{2019Semenlaser,2019Agnesi} or via wire-bounding. We perform a complete demonstration of polarization-encoding MDI-QKD and distill finite-key secret rates higher than previous experiment. This work paves the way for low-cost, wafer-scale manufactured MDI-QKD system, and represents a key step towards building quantum network with untrusted relays\cite{2013Frohli,hughes2013network,wengerowsky2018entanglement}.

\noindent \emph{Notes added}: When we prepare the manuscript, we become aware of a related work in ref.\cite{2019Seme}. Our work uses polarization encoding and low-cost Si substrate at a clock rate of 1.25 GHz, with a careful finite-key consideration and an implementation of random modulations of decoy states and polarization qubits. Ref.\cite{2019Seme} uses time-bin encoding and InP substrate at a clock rate of 0.25 GHz, without the implementation of random modulations and the finite-key consideration, but it integrates the lasers on chip.

\begin{methods}

\subsection{Source}
 Each user consists of two gain-switched lasers of which the wavelength are stabilized using tunable temperature controllers. The master laser and the slave laser are individually driven by 500-ps and 200-ps square-wave pulses with a repetition rate of 1.25 GHz. An electrical delay in steps of 1 ps allows a perfect temporal overlap of two lasers. The seeding photons are injected into the cavity of the slave laser via a circulator. With the seeding photons, the slave lasers generate low-jitter phase-randomized light pulses with a pulse width of $\sim$100~ps. The generated pulses pass through a filter with a bandwidth of 10 GHz to reduce frequency chirp. To test the visibility of the setup, we perform a two-photon interference experiment. The photon count rate is attenuated to $\sim3.5~\text{MHz}$ per detector. Data is collected for 100 s with a coincidence time window of 600 ps, resulting in a visibility of 0.484.

\subsection{Si transmitter chip}
The generated pulses are coupled into a Si photonic transmitter chip  which integrates together an intensity modulator,  and polarization modulator, and variable optical attenuator. The intensity modulator is realized by the first Mach-Zehnder interferometer. By applying  multi-level RF signal to the CDMs, the intensities are randomly modulated   according to four different intensity choices ($\mu$ , $\nu$, $\omega$, $0$). The intensity modulator provides a static extinction ratio (ER) of $\sim$30 dB and a dynamic ER of $\sim$ 20 dB. To reduce the QBER penalty from the ER, we use an external $\text{LiNbO}_3$ intensity modulator to enhance the ER of vacuum state.

The next components is the VOA, consisting of a PIN diode for current injection across-section of the waveguide and being used to attenuated the pulses to single-photon levels.  In our chip, we have three cascade connected VOAs and each VOA provide a $\sim$38 dB dynamic range. By applying differential $dc$-biased voltage to the PIN, the max attenuation is up to $\sim$110~dB. The output of VOA is connected to the polarization modulator (POL) which is realized by combining an inner MZI with two external CDMs ending in polarization rotator combiner (PRC). The inner MZI controls the amplitude ratio between the two arms of the PRC by $dc$-voltage biasing the TOMs in the first inner MZI. Then the PRC converts the transverse-electric polarization in one of its arm into the transverse-magnetic polarization, which is recombined with the light from the other arm at the output. The relative phases of the two arms are modulated by the CDMs in one of arms. Therefore, we obtain the four states required by the protocol as follow. First, we create the polarization $(\left| H \right\rangle  + \left| V \right\rangle )/\sqrt 2 $ by $dc$-voltage biasing the TOMs in the inner MZI. Then, by applying different levels of RF signal on the CDM in one arm of the PRC,  a $\theta  \in \{ 0,{\pi  \mathord{\left/
  		{\vphantom {\pi  2}} \right.
  		\kern-\nulldelimiterspace} 2},\pi ,{{3\pi } \mathord{\left/
  		{\vphantom {{3\pi } 2}} \right.
  		\kern-\nulldelimiterspace} 2}\}$ phase shift is imposed to one of the arm of PRC, creating the one of polarization states $\left| \psi  \right\rangle  =( \left| H \right\rangle  + {e^{i\theta }}\left| V \right\rangle)/\sqrt{2}$.

We characterize the produced state by using the measurement device in Charlie. The EPCs are adjusted so that each PBS is aligned to rectilinear and diagonal bases, respectively. With RF voltages between 0 and 7.5 V, we obtain an average polarization ER of $\sim$23 dB which is sufficient for a low-error MDI-QKD operation.

\subsection{Detection}
The Bell state measurement devices are located in Charlie. The synchronization clock is electrically distributed with a tunable time delay in steps of 1 ps. This enables Alice, Bob and Charlie to electrically compensate any temporal drifts. The projection results are detected with four SNSPDs. The SNSPDs is cooled down to 2.1 K and with an detection efficiency of $\sim$53\%, dead time of $\sim$40 ns, time jitter of $\sim$70 ps and dark counts $\sim$50 Hz. Since the system has a GHz repetition rate, which requires that the SNSPD can tolerate a peak counting rate of more than 5 MHz. We solve it by inserting a 50 ohm shunt resistor between the $dc$ arm of the bias tee and the ground at room temperature. This improved electrical configuration can prevent the detector from latching at a higher count rate without scarifying the detection efficiency. The detection events are recorded by a high speed time tagger with a max data transfer rate up to 65 MHz. The time coincidence time window is set to 600 ps which is an optimal trade-off between the detection efficiency and the error rate of $X$ basis.

\end{methods}

\paragraph{Acknowledgments}

We thank Jianhong Liu, Pan Gong, Yan-Lin Tang, Wenyuan (Mike) Wang for enlightening discussions. This work was supported by the National Key Research and Development (R\&D) Plan of China (under Grants No. 2018YFB0504300 and 2017YFA0304000), the National Natural Science Foundation of China (under Grants No. 61771443 and No. 61705048), the Anhui Initiative in Quantum Information Technologies and the Chinese Academy of Sciences.

\bibliographystyle{naturemag_noURL}

\newpage
\clearpage

\begin{table*}[]
	\small
	\centering
	\caption{Comparison of state-of-the-art MDI-QKD experiments }
	\begin{tabular}{l @{\hspace{0.1cm}} c @{\hspace{0.3cm}} c @{\hspace{0.3cm}} c @{\hspace{0.3cm}} c}
		Reference & Clock rate (MHz) & Channel loss (dB) & Secret key rate (bps)     & Finite-key                                        \\
		\hline  \hline
		Tang et al.\cite{2016zhiyuan}    & 10            & 2.0                & 25$^\star$                         & ${10^{ - 3}}$    \\
		\hline
		Tang et al.\cite{2014yanlin}    & 75            & 9.9                & 67                         & ${10^{ - 9}}$   \\
		\hline
		Valivarthi et al.\cite{2017vali}    & 20            & 16.0                & 100                         & Asymptotic    \\
		\hline
		Yin et al.\cite{2016Yin} & 75             & 19.5               & 1380  & ${10^{ - 10}}$                                
\\
		\hline
		Comandar et al.\cite{2016Comandar} & 1000             & 20.4               & 4567$^\dagger$  & ${10^{ - 10}}$                             \\
		\hline
		\multirow{3}{*}{This work}
		&   & 20.4               &     6172$^{\ddagger}$                      &  ${10^{ - 10}}$     \\
		& 1250  & 28.0               &     268                       &  ${10^{ - 10}}$   \\
		&  & 36.0               &     31                       &  ${10^{ - 10}}$   \\
		\hline  \hline
		\scriptsize
	  $^\star$With encoding flaws\\
	  \scriptsize
		$^\dagger$No random state/decoy modulations\\
		\scriptsize
		$^{\ddagger}$Simulated by experimental parameters

	\end{tabular}

	\label{Comparison}
\end{table*}

\clearpage
\newpage

\section*{SUPPLEMENTARY MATERIALS} 

\paragraph{Four-intensity decoy-state MDI-QKD.}
We realize four-intensity  protocol proposed by Zhou et al.\cite{2016Zhou}. In this protocol,
Alice and Bob each randomly prepares a weak coherent pulse either in the $Z = \{ \left| 0 \right\rangle ,\left| 1 \right\rangle \} $ basis, or in the $X = \{ \left|  +  \right\rangle ,\left|  -  \right\rangle \} $ basis, where $\left| 0 \right\rangle  = {{(\left| H \right\rangle  + \left| V \right\rangle )} \mathord{\left/
		{\vphantom {{(\left| H \right\rangle  + \left| V \right\rangle )} {\sqrt 2 }}} \right.
		\kern-\nulldelimiterspace} {\sqrt 2 }}$, $\left| 1 \right\rangle  = {{(\left| H \right\rangle  - \left| V \right\rangle )} \mathord{\left/
		{\vphantom {{(\left| H \right\rangle  - \left| V \right\rangle )} {\sqrt 2 }}} \right.
		\kern-\nulldelimiterspace} {\sqrt 2 }}$,  $\left|  +  \right\rangle  = ({{\left| H \right\rangle  + {e^{i\frac{\pi }{2}}}\left| V \right\rangle } \mathord{\left/
		{\vphantom {{\left| H \right\rangle  + {e^{i\frac{\pi }{2}}}\left| V \right\rangle } {\sqrt 2 }}} \right.
		\kern-\nulldelimiterspace} {\sqrt 2 }}$ and $\left|  -  \right\rangle  = ({{\left| H \right\rangle  + {e^{i\frac{{3\pi }}{2}}}\left| V \right\rangle } \mathord{\left/
		{\vphantom {{\left| H \right\rangle  + {e^{i\frac{{3\pi }}{2}}}\left| V \right\rangle } {\sqrt 2 }}} \right.
		\kern-\nulldelimiterspace} {\sqrt 2 }}$. There are three intensities $\{\mu,~\nu,~\omega\}$ in the $X$ basis for decoy-state analysis and one signal intensity $\{s\}$ in the $Z$ basis for secret key generation. Including the probabilities $P$ for each intensity, Alice and Bob use the same group of 6 parameters $\left[s, \mu, \nu, P_{s}, P_{\mu}, P_{\nu}\right]$. Here we perform a full optimization of parameters\cite{2014Xu}. For statistical fluctuations, we use the \emph{joint constrains} where the same observables are combined and treated together, as proposed in ref.\cite{2016Zhou}. This can produce a higher key rate than independent constrains. Finally, the secret key is extracted using the formula,
\begin{equation}
\begin{array}{r}{R=P_{s_{\mathrm{A}}} P_{s_{\mathrm{B}}}\left\{\left(s_{\mathrm{A}} e^{-s_{\mathrm{A}}}\right)\left(s_{\mathrm{B}} e^{-s_{\mathrm{B}}}\right) Y_{11}^{X, L}\left[1-h\left(e_{11}^{X, U}\right)\right]\right.}  {\left.-f_{e} Q_{s s}^{Z} h\left(E_{s s}^{Z}\right)\right\}}\end{array},
\end{equation}
where $Q_{s s}^{Z}$ and $E_{s s}^{Z}$ are the gain and quantum bit error rate $\mathrm{QBER}$ in the $\mathrm{Z}$ (signal) basis, $P_{s_{\mathrm{A}}} ~( P_{s_{\mathrm{B}}})$ is the probability of signal state for Alice (Bob),  $Y_{11}^{X, L}$ and $e_{11}^{X, U}$ are the lower bound of single-photon yield and the upper bound of QBER, estimated by the decoy state
statistics in the $\mathrm{X}$ basis, $h$ is the binary entropy function, and $f_{e}$ is the error-correction efficiency, which is set to 1.16.

\paragraph{Experimental details.}

\subsection{Source.}
\begin{figure}[!hbt]
	\centering
	\includegraphics[width=0.6\linewidth]{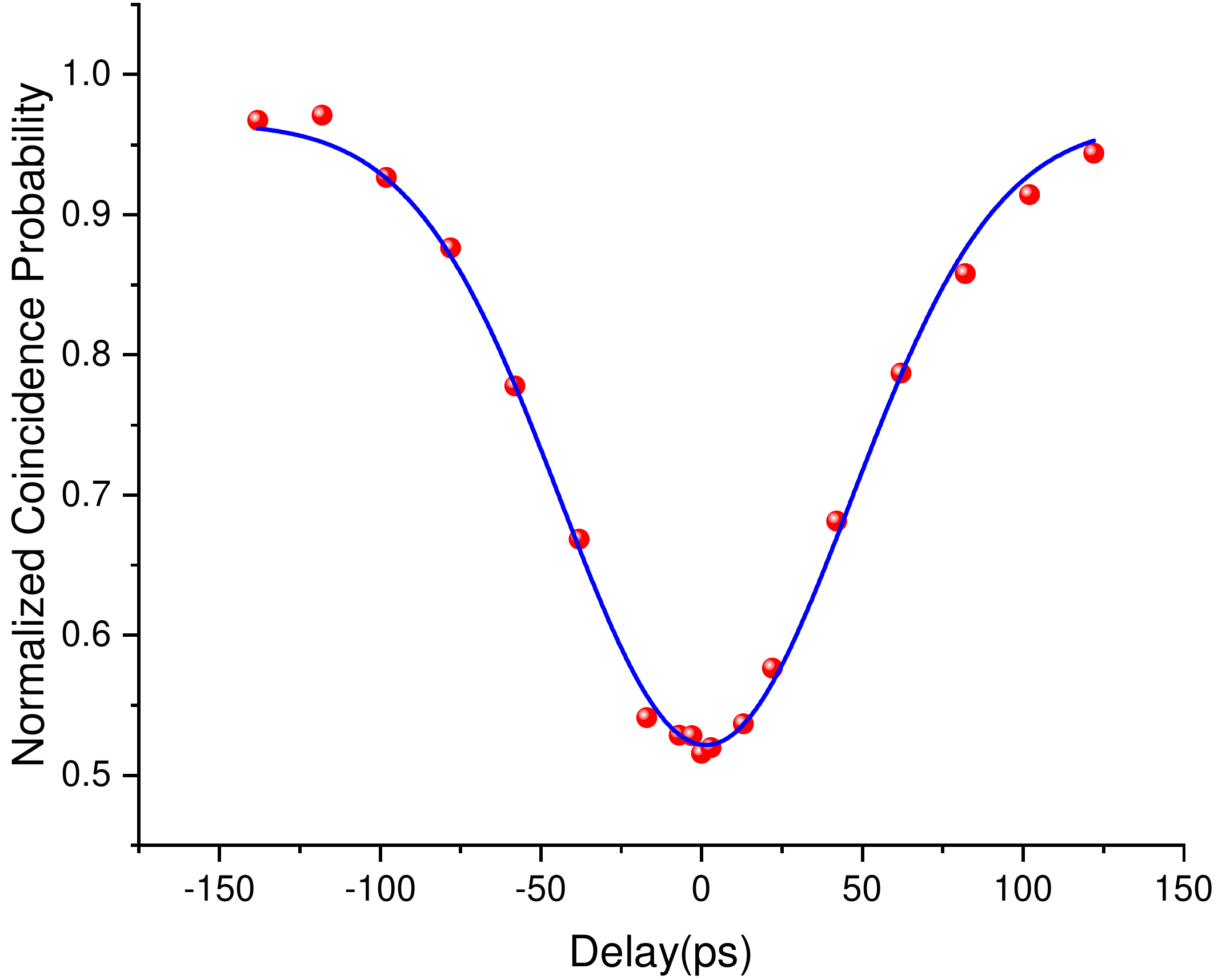}
	\caption{Hong-Ou-Mandel interference between two gain-switch lasers using laser seeding technology. A dip with visibility 48.4\% is obtained.}\label{HOM}
\end{figure}

In our setup, Alice and Bob each generates the 1.25-GHz laser pulses using laser seeding technology. To test the visibility of the setup, we perform a two-photon Hong-Ou-Mandel interference experiment. The electronic polarisation controllers (EPCs) in Charlie are adjusted so that each polarised beam splitter (PBS) is aligned to rectilinear basis. The photon count rate is attenuated to $\sim3.5~\text{MHz}$ per detector. Data is collected for 100 s with a coincidence time window of 600 ps. As shown in Fig. \ref{HOM}, we obtain a visibility of 48.4\%.

\subsection{Si chip transmitter.}
\begin{figure}[!hbt]
	\centering
	\includegraphics[width=0.6\linewidth]{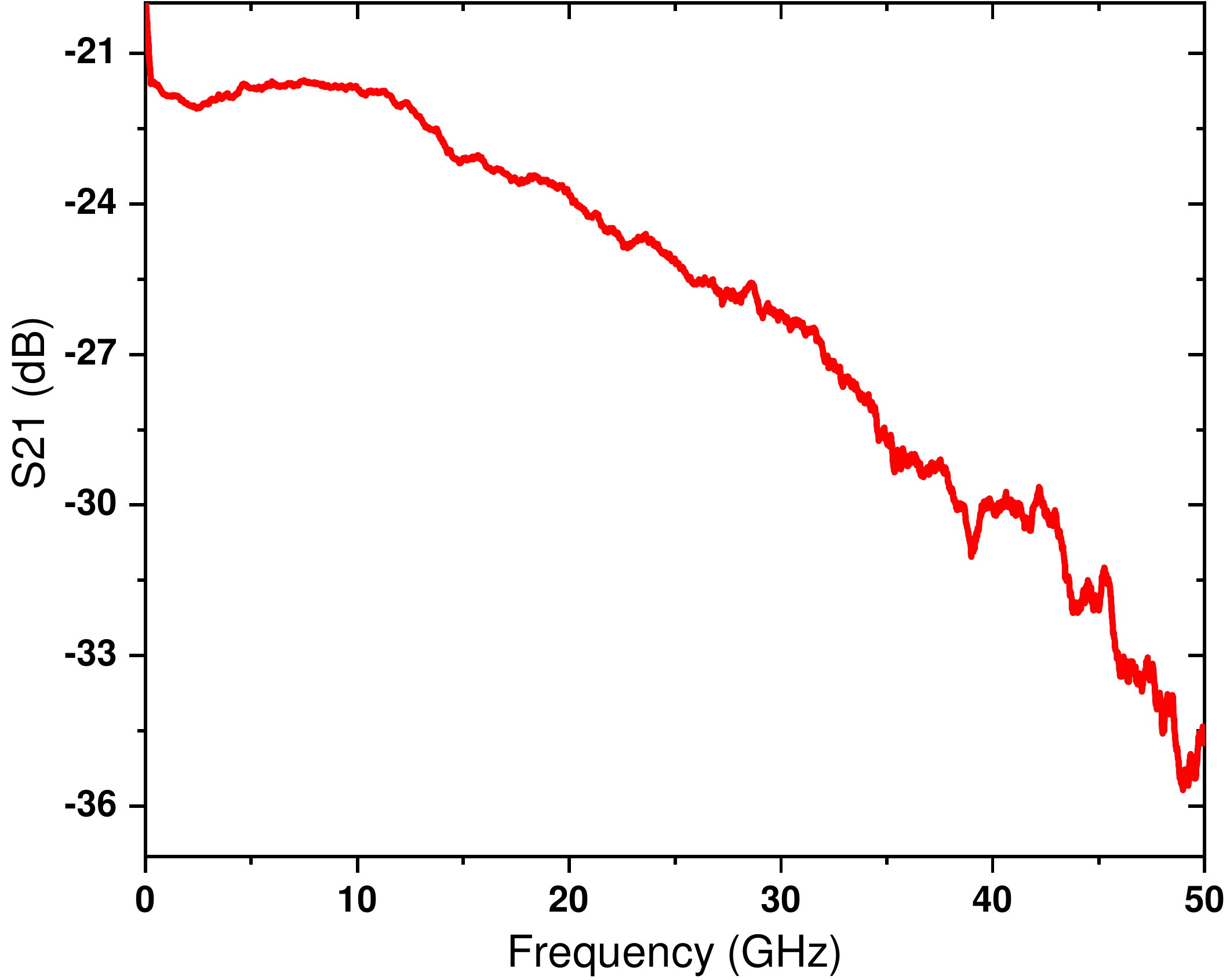}
	\caption{S21 curve of CDM.  A 3-dB bandwidth of $\sim$21 GHz is achieved.  }\label{S21}
\end{figure}

A CDM, acting as a phase modulator, is a key component in our chip. The bandwidth of the CDM must be subtly estimated since it has a crucial role on the performance of intensity modulators (IMs) and polarization modulators (Pols). We measure the bandwidth of the CDM by using a vector network analyzer. As shown in Fig. \ref{S21}, we achieve an 3-dB bandwidth of $\sim$21 GHz.

\begin{figure}[!hbt]
	\centering
	\includegraphics[width=0.6\linewidth]{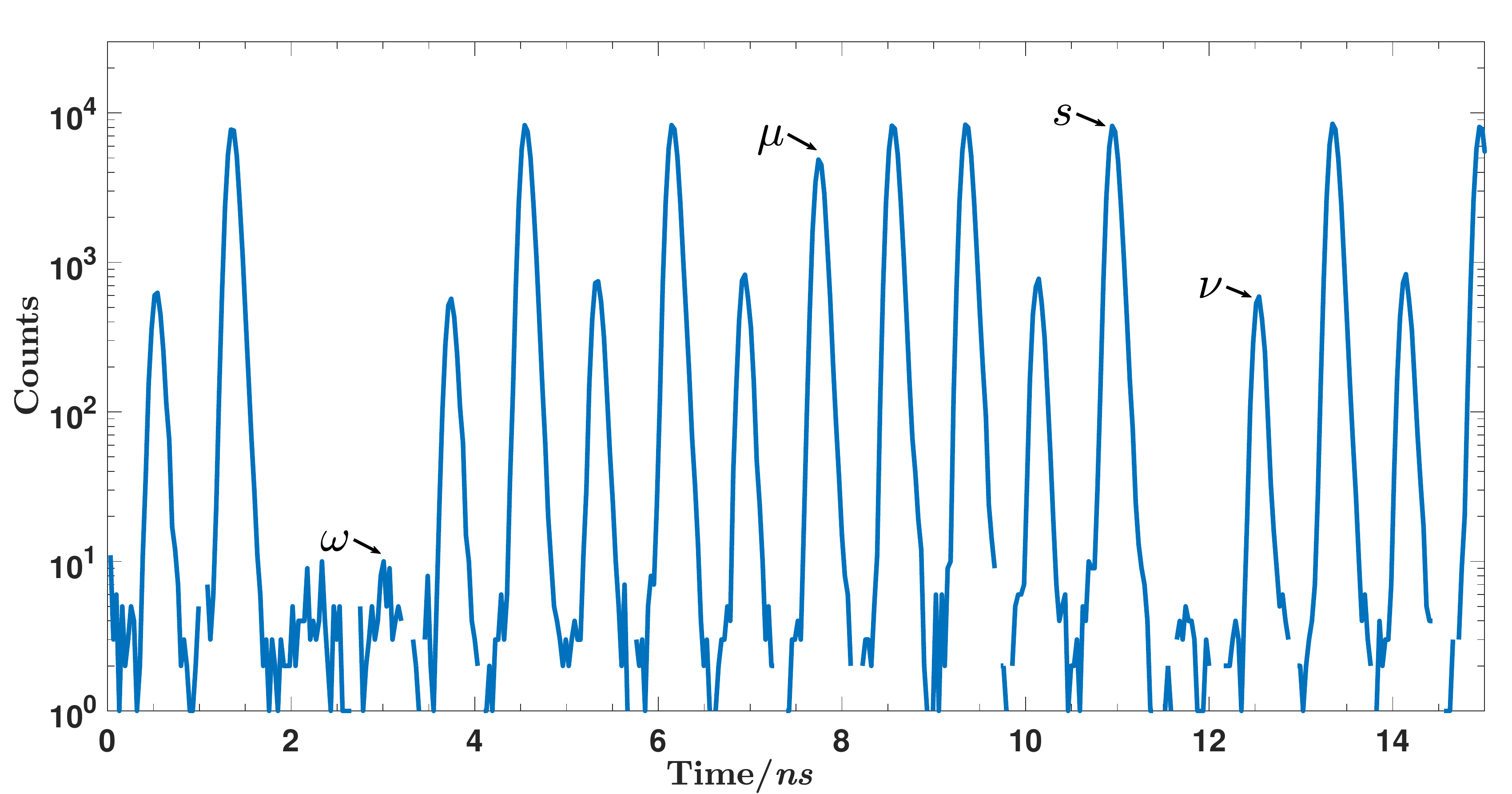}
	\caption{ Histogram measurement of different intensities using the chip. Four varying intensities can be used for signal ($s$) and three decoy states ($\mu,~\nu,~\omega$).}\label{IM}
\end{figure}

Alice and Bob each needs to prepare laser pulses in four intensities which are realized by the intensity modulator on the chip. In experimental characterization, a static extinction ratio (ER) of 29.7 dB is achieved with an applied $dc$-voltage of 0.97 V. In the dynamic ER test, we first trigger the IMs with 625-MHz RF signals and get an ER of 21.5 dB. Then, we enhance the repetition rate of RF signals to 1.25-GHz, and an ER of 19.8 dB is obtained. The decline of the ERs with rising repetition rate is caused by electronic jitter. To further test its performance, we randomly drive the IM with four different RF voltages at a repetition rate of 1.25 GHz, which are generated by our home-made FPGA control board (See Fig. \ref{FPGA}). As shown in Fig. \ref{IM}, the four voltage levels produce four intensities which can be used for signal state ($s$) and three decoy states ($\mu,~\nu,~\omega$).

\begin{figure}[!hbt]
	\centering
	\includegraphics[width=0.6\linewidth]{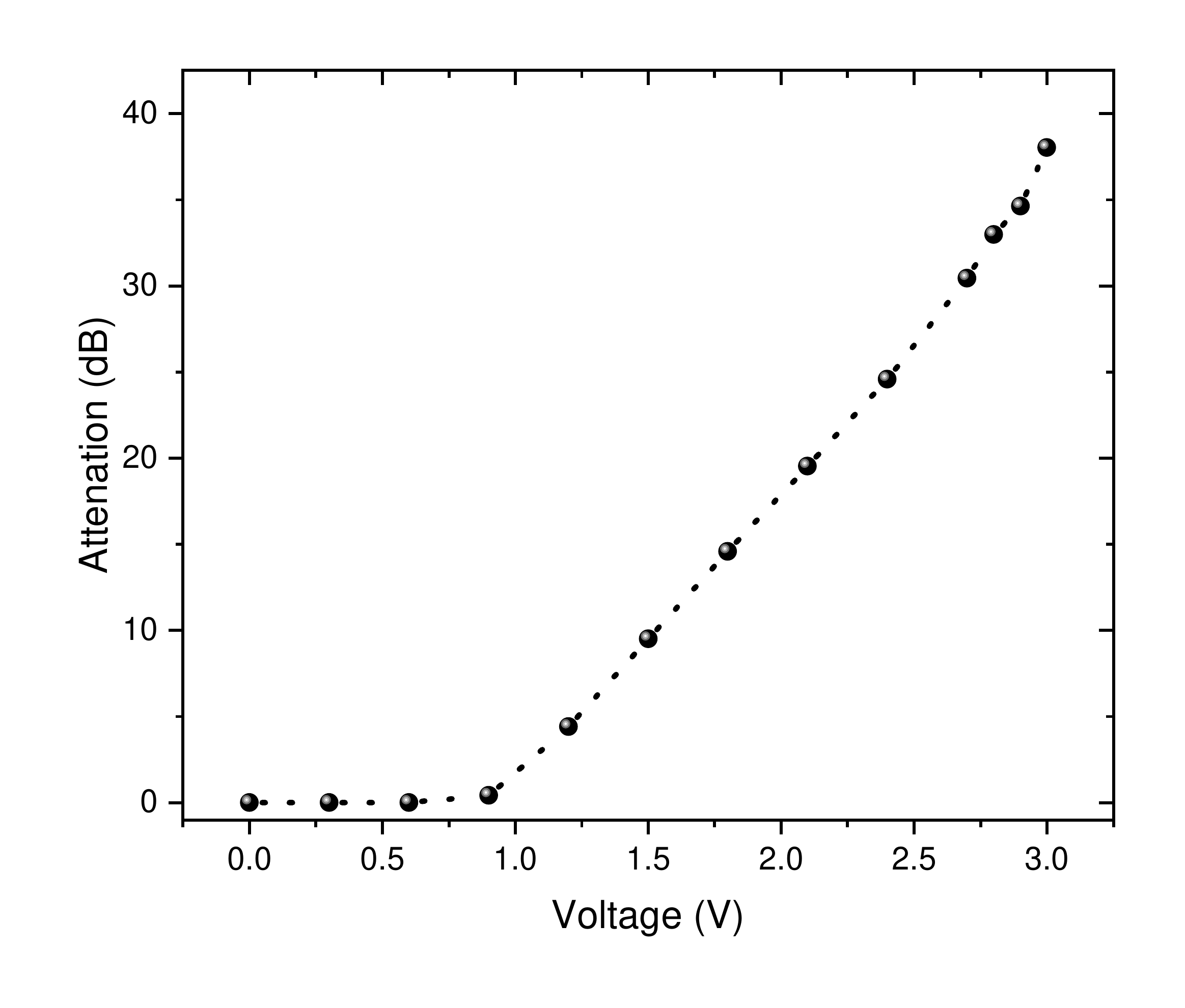}
	\caption{ Measured attenuation versus $dc$-voltage. The maximum attenuation is 38~dB with applying a voltage of 3 V.}\label{VOA}
\end{figure}

A variable optical attenuation (VOA) is realized by using a p-i-n (PIN) diode structure where a PIN diode is designed for current injection across-section of the waveguide. In our chip, we have three cascade connected VOAs. Figure \ref{VOA} shows the tuning ranges of one of the VOAs, which could provide 38.0 dB of attenuation. By applying differential $dc$-biased voltage to the PIN on each VOA, the max attenuation is up to $\sim$110~dB which is sufficient to attenuate laser pluses to single-photon levels.

A polarization modulator  is used to prepare the four states in conjugate bases, e. g., the key generation basis $Z=\{|0\rangle,|1\rangle\}$, where $\left| 0 \right\rangle=\left| H \right\rangle  + \left| V \right\rangle )/\sqrt 2$, $\left| 1 \right\rangle=\left| H \right\rangle  - \left| V \right\rangle )/\sqrt 2$, and the basis $X=\{|+\rangle,|-\rangle\}$, where $\left| + \right\rangle=\left| H \right\rangle  + e^{i\pi/2 }\left| V \right\rangle )/\sqrt 2$, $\left| - \right\rangle=\left| H \right\rangle  - e^{i3\pi/2 }\left| V \right\rangle )/\sqrt 2$.

\begin{figure}[!hbt]
	\centering
	\includegraphics[width=0.4\linewidth]{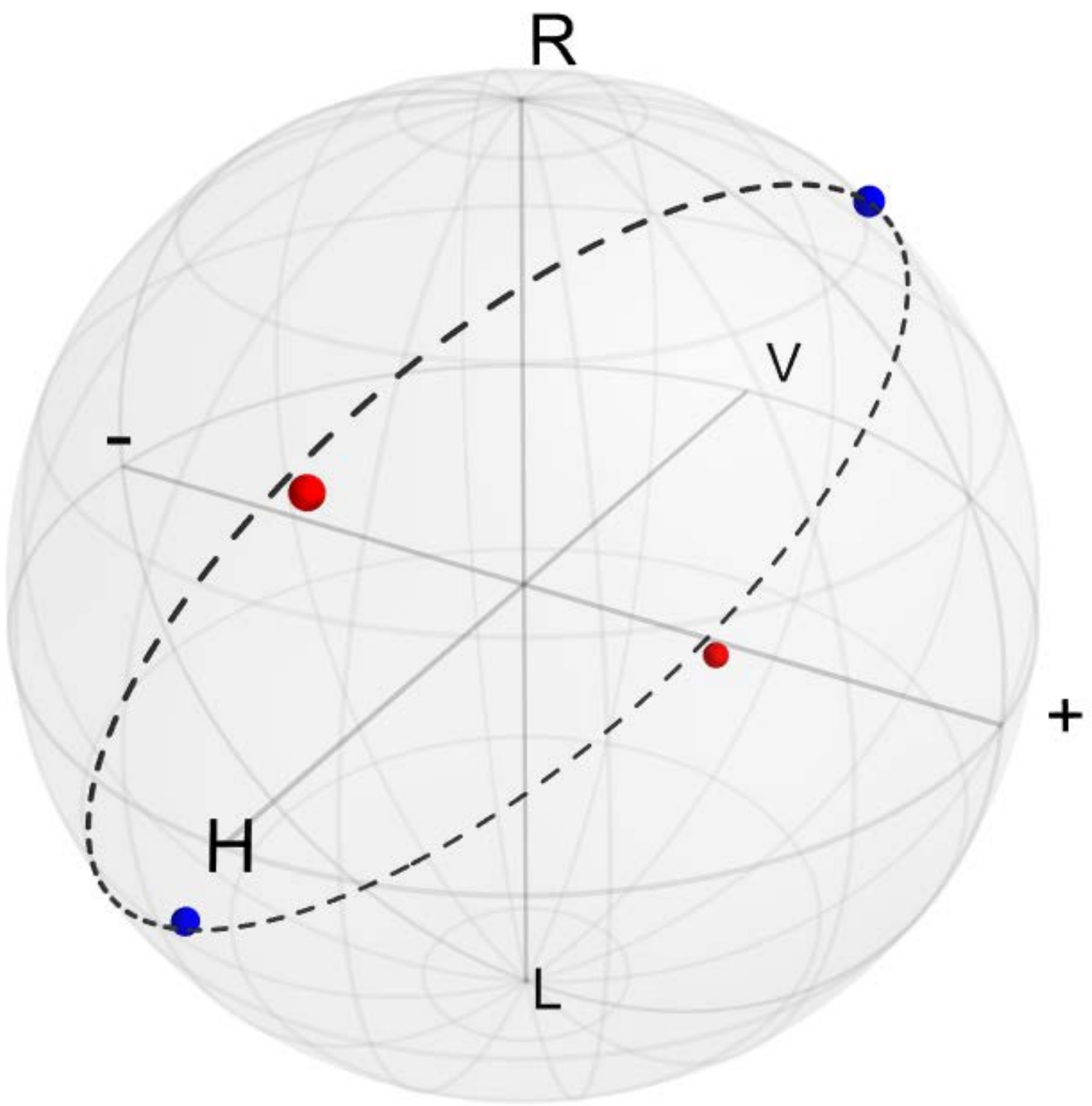}
	\caption{ Measured polarization states on Bloch sphere. The dots represent the produced states by the polarization modulator in chip. }\label{POL}
\end{figure}

We first characterize the produced states by using a polarimeter system (Thorlabs PAX1000). As shown in Fig. \ref{POL},  with appropriate RF signals, we prepare four polarized states in conjugate bases,  exhibiting a well  ER ($\sim$26 dB) and a high degree of polarization ($\sim$0.993). Then, we measure the produced states with the measurement device in Charlie. The EPCs are adjusted so that each PBS is aligned to rectilinear and diagonal bases, respectively. With RF voltages between 0 and 7.5 V, we obtain an average polarization ER of $\sim$23 dB which is sufficient for a low-error MDI-QKD operation.

\subsection{Polarization alignment.}
\begin{figure}[!hbt]
	\centering
	\includegraphics[width=1\linewidth]{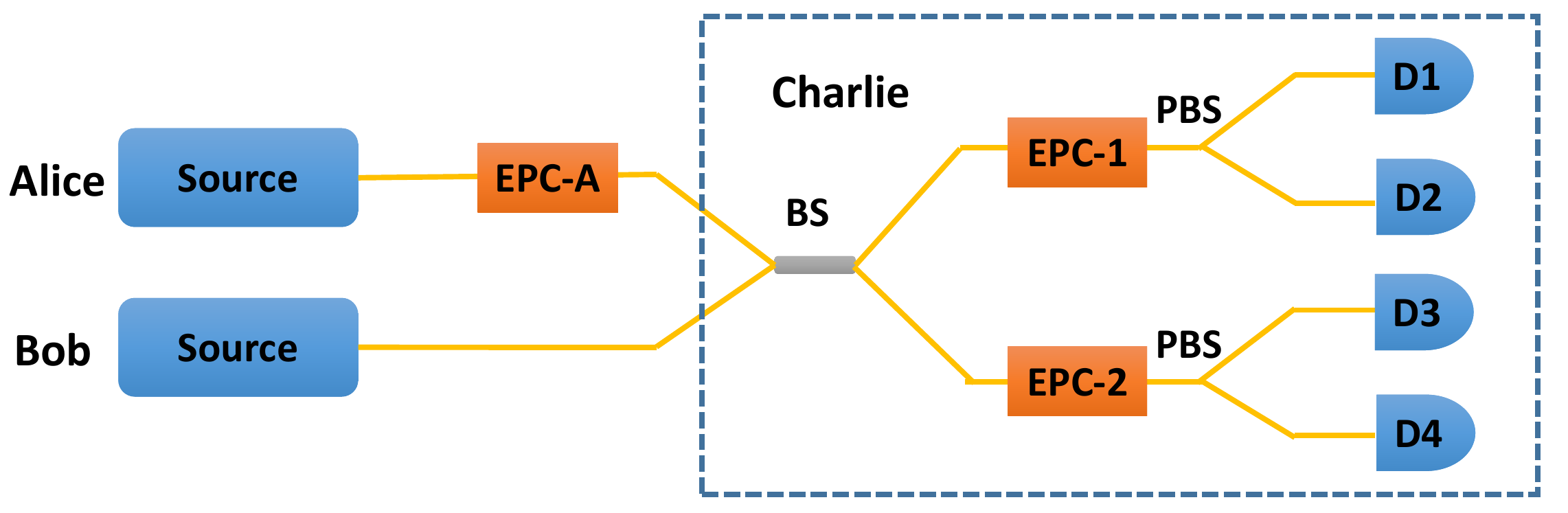}
	\caption{ Schematic of our polarisation alignment system. EPC-A, EPC-1, EPC-2: electronic polarisation controller; BS: beam splitter; PBS: polarisation beamsplitter; D1, D2, D3, D4: superconducting nanowire detector. }\label{EPC}
\end{figure}

\begin{figure}[!hbt]
\centering
\subfigure[]{\label{results1}
	\begin{minipage}[t]{0.3\linewidth}
		\centering
		\includegraphics[width=1\linewidth]{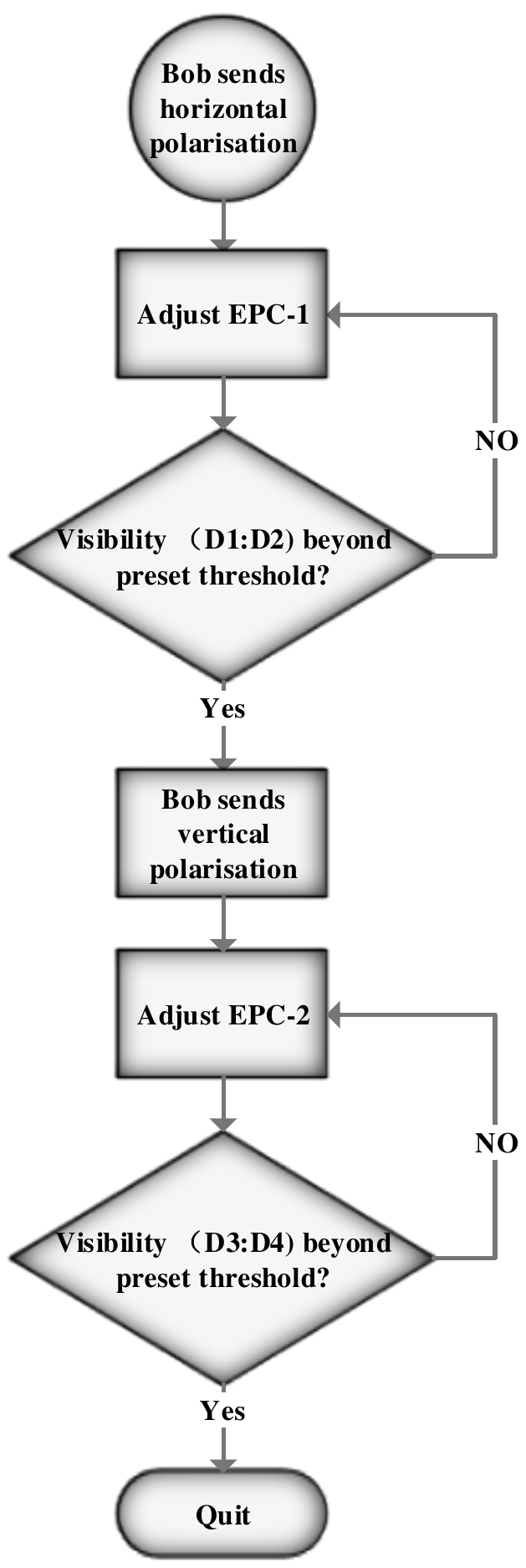}
\end{minipage}}
\subfigure[]{\label{results2}
	\begin{minipage}[t]{0.3\linewidth}
		\centering
		\includegraphics[width=1.2\linewidth]{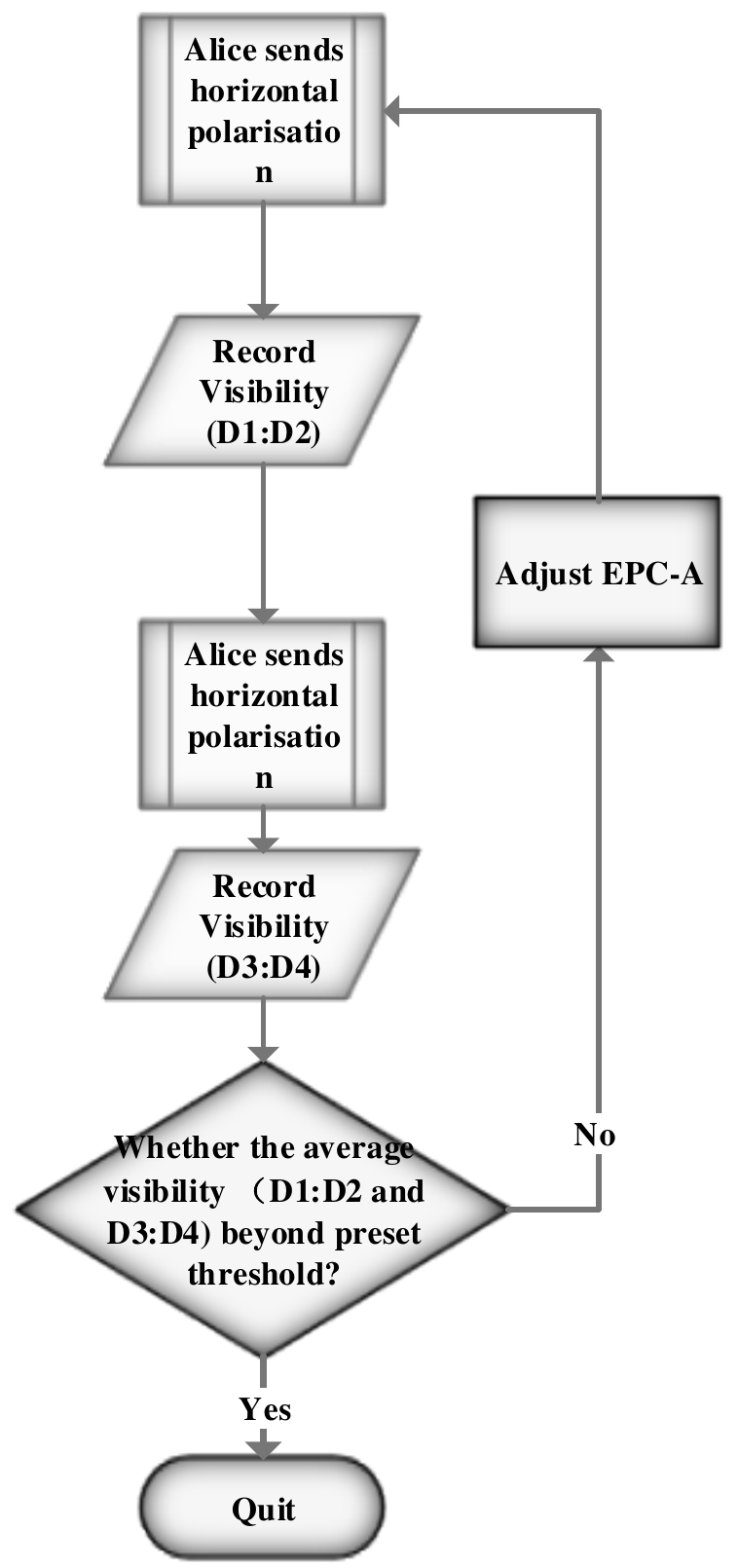}
	\end{minipage}}
		\subfigure[]{\label{results3}
			\begin{minipage}[t]{0.3\linewidth}
				\centering
				\includegraphics[width=1\linewidth]{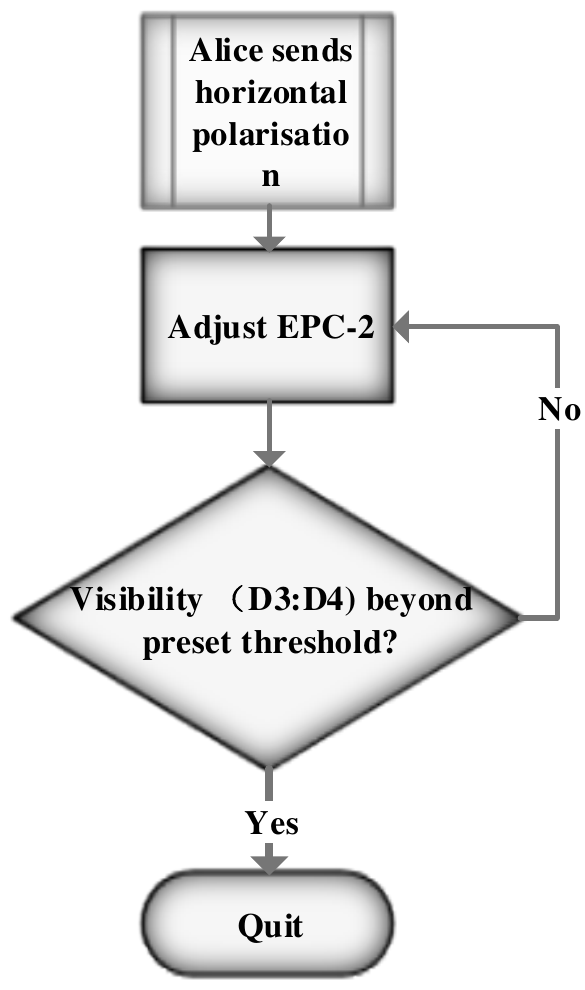}
\end{minipage}}
\caption{Flowchart of polarization alignment.}\label{step}
\end{figure}

Alice and Bob need to share a common polarisation reference as well as compensate the polarisation drift in the quantum channel. Here, we develop an automatic polarisation alignment method which rapidly calibrates the polarisation reference. Figure \ref{EPC} shows the schematic of our alignment system, which is extracted from the setup in the main text. The system can be summarized in the following three steps, which can be realized by following the flowchart in Fig. \ref{step}.

Step 1: Bob and Charlie share a common reference by adjusting EPC-1 and EPC-2, following the flowchart in Fig.~\ref{results1}.

Step 2: Alice aligns her bases by adjusting EPC-A, following the flowchart in Fig. \ref{results2}.

Step 3: Charlie align one of PBS to the Z-basis by adjusting EPC-2, following the flowchart in Fig.~\ref{results3}.

At last, Alice, Bob and Charlie automatically share a common polarization reference at a short time.

\subsection{Electronic control board.} \label{Fpga}
\begin{figure}[!hbt]
	\centering
	\includegraphics[width=0.5\linewidth]{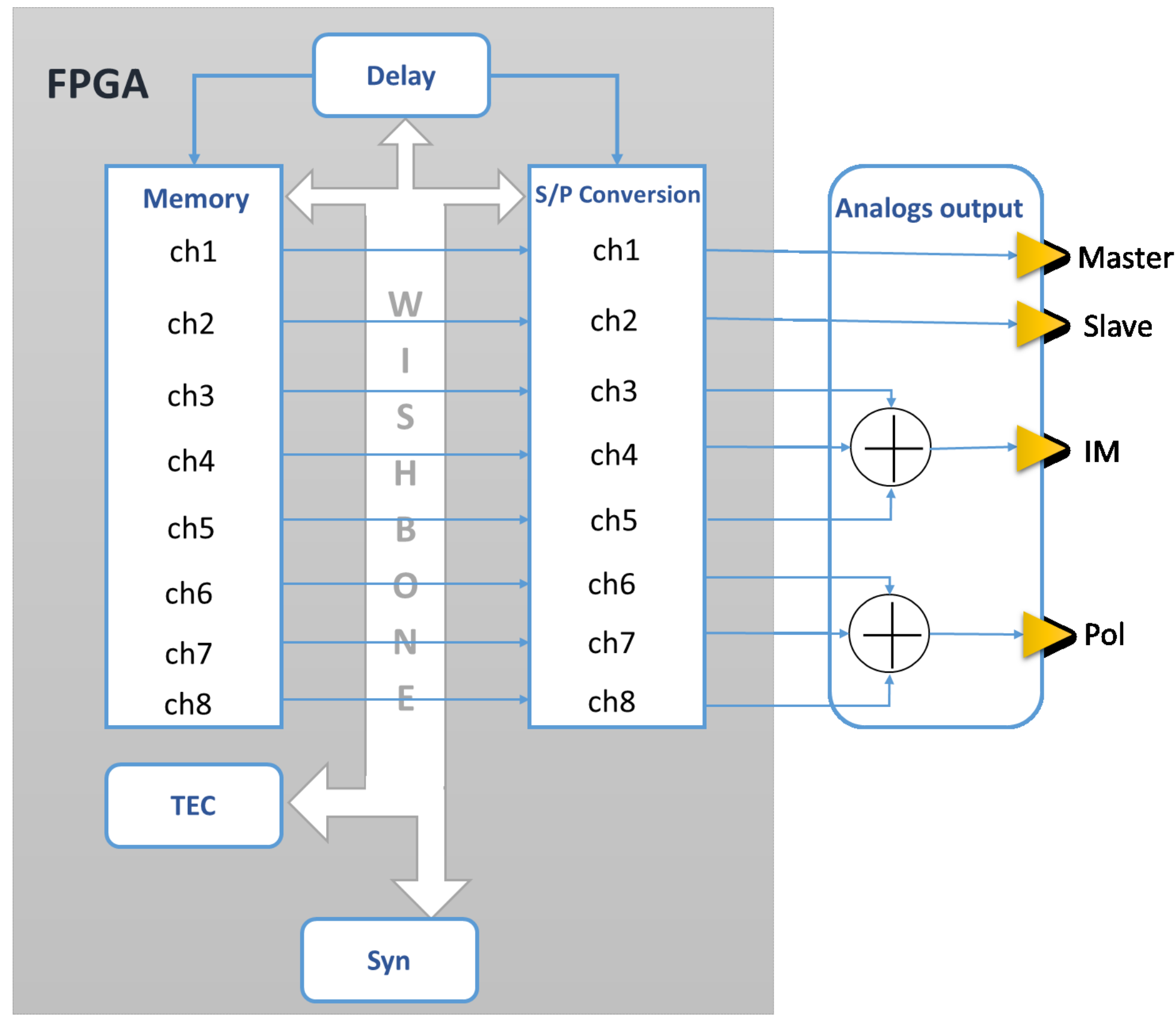}
	\caption{ Architecture of the electronics board. The dash box denotes a FPGA, which can be divided into several modules. }\label{FPGA}
\end{figure}

Figure \ref{FPGA} shows the architecture of the FPGA board, which mainly consists of memory module, serial/parallel conversion module (S/P conversion), delay module,  thermoelectric cooler (TEC) module, synchronization module (Syn) and analogs output module. All the modules (except the analogs output module) are implemented on a Xilinx Kintex-7 FPGA. 8 digital channels are able to generate signals with 10 GHz sampling rate. Among them, two sets of channels (3,4,5 and 6,7,8) are synthesized to output multi-level signals for driving IM and Pol, respectively. After being amplified to 8 V, the analog outputs reach a bandwidth of 2.5 GHz thanks to the careful impedance matching.

The memory module provides 64 kb for 8 digital channels each to storage waveform. The S/P conversion module is used to convert the the low-speed serial data to high-speed parallel data. With this conversion, the data transmission rate reaches 12.5 Gbps, which meets the requirement for transiting data stream in parallel channels simultaneously. The delay module can adjust the delay of each channel. It has a resolution of 1~ps which provides a perfect overlap of the laser pulses and accurate intensity/polarisation modulations. The analogs output is used to amplify the signals from the FPGA. Since IMs and Pols require four amplitude voltages for modulation, there is a 50-GHz pulse combiner to create four amplitude voltages. The TEC module generates all signals for thermoelectric controls, such as the lasers. The Syn module is designed to create up to 4 delayed output pulse sequences precisely synchronized to internal or external clock. Using this pulse sequences, we can synchronize all stations in our setup.

\paragraph{Detailed experimental results}
The detailed experimental results are listed in Table \ref{result}.

\newpage


\begin{figure}[!hbt]
	\centering
	\includegraphics[width=0.5\linewidth]{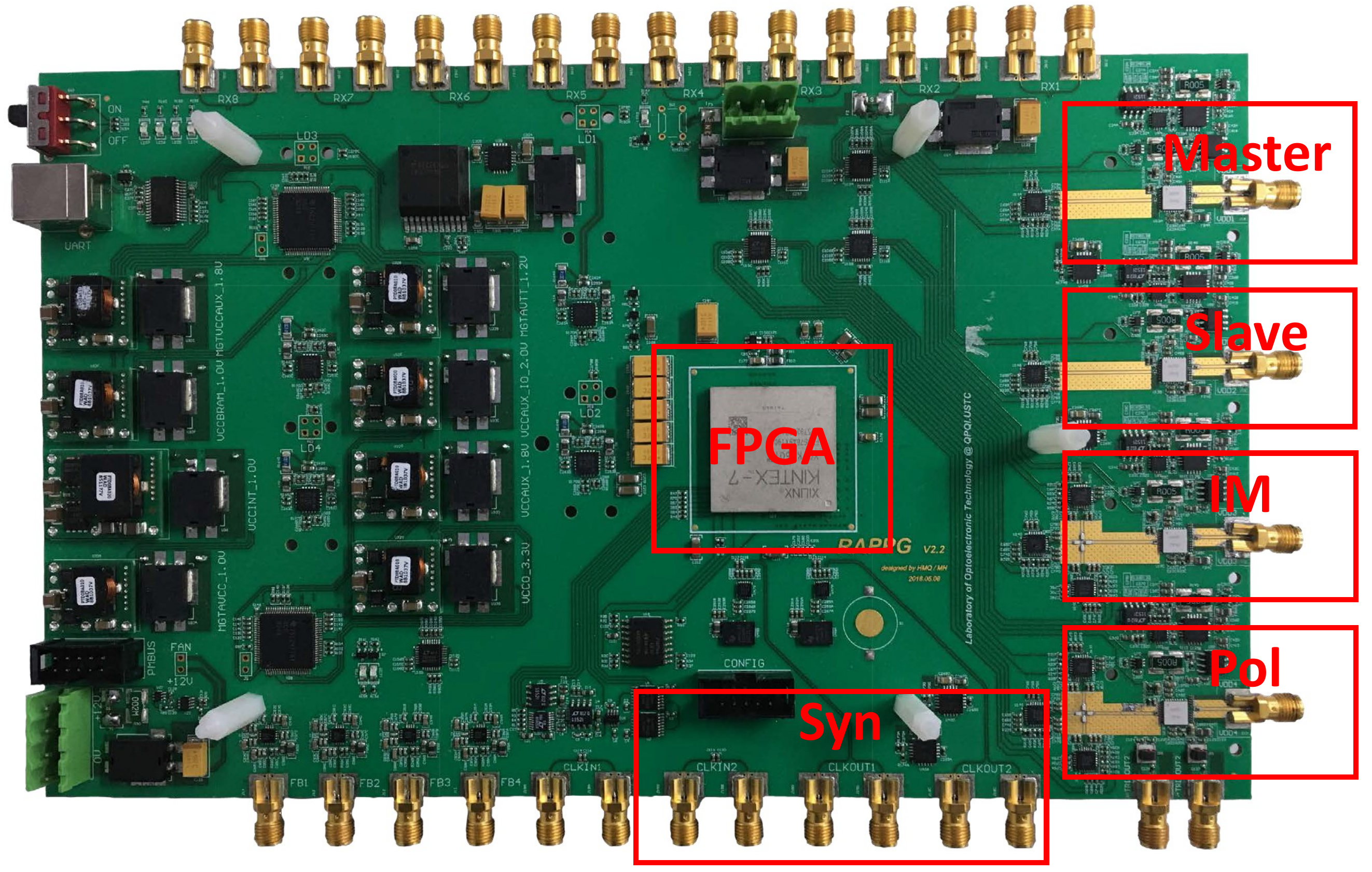}
	\caption{ Image of the electronics board. }\label{FPGA-1}
\end{figure}

\begin{table}
	\centering
	\caption{Total gains and error gains of Bell state $\left| {{\psi ^ \pm }} \right\rangle  $. The notation $Q_{ij}$ and $E_{ij}$ denotes the total gains and error gains from Alice's source $i$ and Bob's source $j$, respectively. Source $ij$ and source $ji$ are regarded as one complex source which is represented by $ij+ji$.}
	\begin{tabular}{c|cc}
		\hline\hline	
		$\rm Attenuation~(dB)$   &$\rm28.0$&$\rm36.0$\\
		\hline
		$\rm Distance~(km)$  & $\rm 140$ &$\rm 180$ \\
		\hline
		$N$  &$3.0\times 10^{13}$&$4.5\times 10^{13}$\\
		\hline
		$Q_{ss}$& $67610084 $& $6305857  $\\
		$Q_{\mu\mu}$&$258557$
		&$115035$\\
		
		$Q_{\nu\nu}$&$606196$
		
		&$388040$\\
		
		$Q_{\mu0+0\mu}$& $151827$ & $226256  $\\
		
		$Q_{\nu0+0\nu}$& $116967 $& $71521  $\\
		
		$Q_{00}$& $6$& $0  $\\
		\hline
		$E_{ss}$& $1851744$& $178909  $\\
		$E_{\nu\nu}$& $160177$& $104895  $\\
		$E_{\nu0+0\nu}$& $58342$& $35874  $\\
		$E_{00}$& $4$& $0  $\\
		\hline
		$s_{11}$&$8.98\times 10^{-5}$&$2.43\times 10^{-5}$\\
		$e_{11}$&$0.068$&$0.089$\\	
		$\rm Key~rate/pulse$&  $1.29\times 10^{-7}$&     $2.47\times 10^{-8}$\\
		\hline\hline
	\end{tabular}
	\label{result}
\end{table}

\end{document}